\documentclass[prl,twocolumn,showpacs,preprintnumbers,nofootinbib]{revtex4-1}
\usepackage{amsfonts,amsmath,amssymb}
\usepackage[english]{babel}
\usepackage{graphicx}
\usepackage{bm}
\usepackage{braket}
\usepackage[caption=false]{subfig}
\usepackage{xcolor}
\usepackage[hyperindex,breaklinks]{hyperref}
\graphicspath{ {./jaxo/} }

\begin{document}
\preprint{P3H-19-54, TPP19-48, SI-HEP-2019-22, JLAB-THY-19-3123}

\title{\boldmath $\tau \to \mu\mu\mu$ at a rate of one out of $10^{14}$ tau decays?}
\author{Patrick Blackstone} 
\email{pblackst@iu.edu}
\affiliation{Department of Physics, Indiana University, Bloomington, IN 47408, USA}
\affiliation{Center for Exploration of Energy and Matter,
Indiana University, Bloomington, IN 47408, USA}

\author{Matteo Fael}
\email{matteo.fael@kit.edu}
\affiliation{Institut f\"ur Theoretische Teilchenphysik, Karlsruhe Institute of Technology (KIT), 
76131 Karlsruhe, Germany}
\affiliation{Theoretische Physik I, Universit\"at Siegen, 
57068 Siegen, Germany}

\author{Emilie Passemar}
\email{epassema@indiana.edu}
\affiliation{Department of Physics, Indiana University, Bloomington, IN 47408, USA}
\affiliation{Center for Exploration of Energy and Matter,
Indiana University, Bloomington, IN 47408, USA}
\affiliation{Theory Center, Thomas Jefferson National
Accelerator Facility, Newport News, VA 23606, USA}

\begin{abstract} 
\noindent 
We present in a full analytic form the partial widths for the lepton flavour violating decays \mbox{$\mu^\pm \to e^\pm e^+ e^-$} and $\tau^\pm \to \ell^\pm \ell'^{+} \ell'^{-}$, with $\ell,\ell'=\mu,e$, mediated by neutrino oscillations in the one-loop diagrams. Compared to the first result by Petcov in~\cite{Petcov:1976ff}, obtained in the zero momentum limit $\mathcal{P}\ll m_{\nu} \ll M_W$, we retain full dependence on $\mathcal{P}$, the momenta and masses of external particles, and we determine the branching ratios in the physical limit $m_\nu \ll \mathcal{P} \ll M_W$. 
We show that the claim presented in~\cite{Pham:1998fq} that the $\tau \to \ell \ell' \ell'$ branching ratios could be as large as $10^{-14}$, as a consequence of keeping the $\mathcal{P}$ dependence, is flawed. 
We find rates of order $10^{-55}$, even smaller than those obtained in the zero momentum limit, as the latter prediction contains an unphysical logarithmic enhancement. 
\end{abstract} 

\date{\today}
\maketitle

\section{Introduction}
It is reported by several experimental collaborations, e.g.\ by CMS~\cite{CMS:2019sxo}, ATLAS~\cite{Aad:2016wce}, LHCb~\cite{Aaij:2014azz}, BABAR~\cite{Aubert:2003pc,Aubert:2007pw,Aubert:2007kx,Lees:2010ez} and Belle~\cite{Hayasaka:2010np}, that the branching ratios for the charged lepton flavour violating (CLFV) decays $\tau^\pm \to \ell^\pm \ell'^{+} \ell'^{-}$, with $\ell, \ell' = e, \mu$, can be as large as $10^{-14}$ in the Standard Model extended with  either a Dirac or a Majorana mass term for neutrinos. 
This follows from a claim by Pham in~\cite{Pham:1998fq} that for these decays the GIM mechanism~\cite{Glashow:1970gm} produces a suppression of the form $|\sum_i U_{\ell i} U^*_{Li} \log x_i|^2$, where $x_i = m_{\nu i}^2/M_W^2$, $i=1,2,3$, $m_{\nu i}$ and $M_W$ are the masses of the three neutrinos and the $W$ boson, and $U$ is the Pontecorvo-Maki-Nakagawa-Sakata (PMNS) mixing matrix~\cite{Pontecorvo:1967fh,Maki:1962mu}.
The result in~\cite{Pham:1998fq} is in sharp contrast with the first evaluation by Petcov~\cite{Petcov:1976ff}, which showed that these CLFV decays are instead power suppressed  by $\vert\sum_i U_{\ell i} U^*_{L i} \, x_i \log x_i\vert^2$, so that the smallness of the ratios $m_{\nu i}/M_W$ crushes the branching fractions well below~$10^{-54}$, far beyond the sensitivity of any foreseeable experiment.

The calculations in~\cite{Petcov:1976ff} and~\cite{Pham:1998fq} differ as follows. Ref.~\cite{Petcov:1976ff} employed for the evaluation of the one-loop diagrams the zero-momentum-limit (ZML) approximation, which assumes vanishing momenta and masses of the external particles while it retains the dependence on the internal masses of neutrinos and the $W$ boson.
The ZML implicitly assumes the mass scale hierarchy
\begin{equation}
  \mathrm{(ZML)} \quad  \mathcal{P} \ll m_{\nu i} \ll M_W, \notag
\end{equation}
where $\mathcal{P}$ generically stands for any of the external particle momenta and masses, e.g.~$\mathcal{P} \sim m_L$ or $\mathcal{P} \sim m_\ell$. 
This approximation, even if far from the physical situation, allows a substantial simplification of the one-loop integrals, as in this way they depend only on $x_i$.
Ref.~\cite{Pham:1998fq}, on the contrary, argued that once the external momentum dependence is taken into account, the GIM cancellation in $L \to \ell \ell'\ell'$, with $L=\tau$ or $\mu$, becomes actually much milder, with a suppression only of the form $|\sum_i U_{\ell i} U^*_{Li} \log x_i|^2$, which leads to branching ratio values of the order of $10^{-14}$.

If the prediction in~\cite{Pham:1998fq} were true, it would imply, with current values for neutrino mixing angles and mass splittings~\cite{Tanabashi:2018oca}, that the branching ratio of $\mu \to eee$ could reach $10^{-17}$, for a lightest-neutrino mass of the order of $10^{-10}$~eV or smaller. 
This would be just around the corner for the \textsc{Mu3e} experiment currently under development at the Paul Scherrer Institute in Switzerland, which aims to reach a sensitivity of  Br$(\mu \to e e e) \sim 10^{-16}~$\cite{Blondel:2013ia}.
For the tau, the rates would be in the range $10^{-16} - 10^{-13}$, still several orders of magnitude smaller than the current world averages, Br$(\tau \to \ell \ell' \ell') \lesssim 10^{-8}$~\cite{Amhis:2019ckw}, and the expected sensitivity of Belle II, Br$(\tau \to \ell \ell' \ell') \lesssim 10^{-10}$~\cite{Kou:2018nap}, and the HL-LHC, Br$(\tau \to \ell \ell' \ell') \lesssim 10^{-9}$~\cite{Cerri:2018ypt,Fiorendi:2018qwm,Azzi:2019yne}, but more than forty orders of magnitude larger than the prediction of~\cite{Petcov:1976ff}. 

Therefore, the question we address in this letter is if the branching ratios of $L\to \ell \ell'\ell'$ can really change so dramatically once one assumes the physical limit (PL), i.e.\ the hierarchy
\begin{equation}
  (\mathrm{PL}) \quad m_{\nu i} \ll \mathcal{P} \ll M_W, \notag
\end{equation}
instead of the ZML, and keeps full dependence of external momenta and masses in the loop diagrams.

\begin{figure*}[thb]
\centering
  \subfloat[\label{fig:Zpen}]{
    \includegraphics[width=0.5\columnwidth]{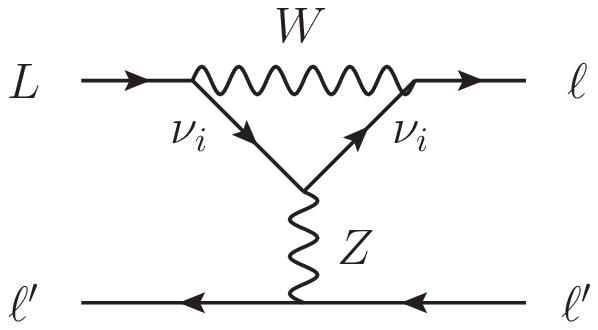}}
  \subfloat[]{
    \includegraphics[width=0.5\columnwidth]{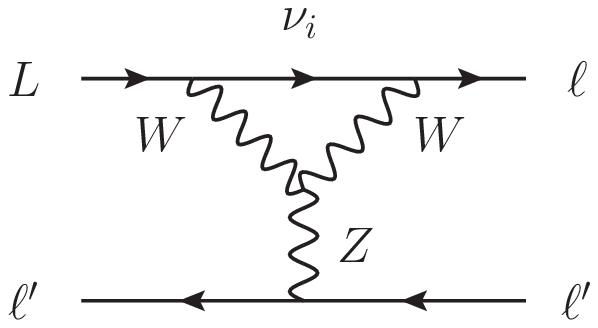}}
  \subfloat[\label{fig:Apen}]{
    \includegraphics[width=0.5\columnwidth]{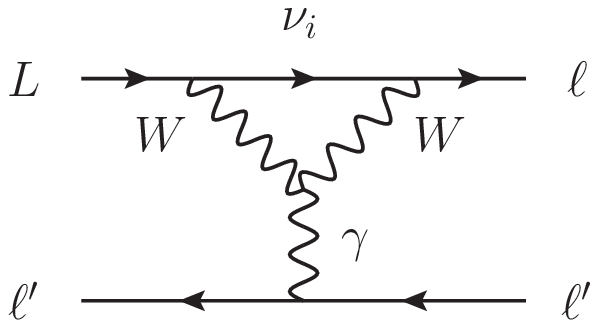}}
  \subfloat[\label{fig:box}]{
    \includegraphics[width=0.5\columnwidth]{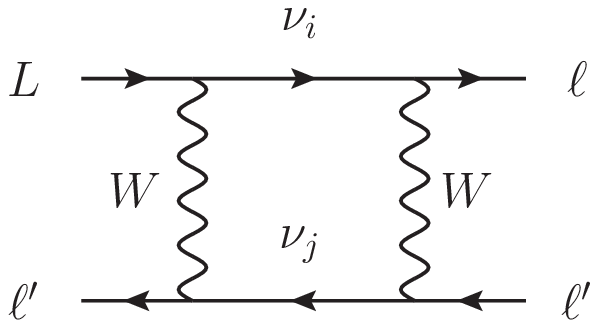}}
    \caption{One-loop diagrams contributing to the CLFV decay $L^- \to \ell^- \ell'^+\ell'^-$ in the unitary gauge: the $Z$ penguins (a,b), the photon penguin (c) and the box (d).
    Wave function corrections (not depicted here) must be included as well.
    }
  \label{fig:fd}
\end{figure*}
In~\cite{Pham:1998fq} it is argued that in the $Z$-penguin diagram shown in Fig.~\ref{fig:Zpen} there are two propagators of nearly massless fermions, which give rise to a $\log x_i$ when the momentum $q$ of the $Z$ boson approaches $q^2 = 0$. This argument is supported by a computation of the $Z$-penguin as an expansion in $q^2/M_W^2$ taking the form $f(q^2,x_i) = f_0(x_i) + (q^2/M_W^2) f_1(x_i) + \dots$. By noticing that $f_0 \sim x_i \log x_i$ is suppressed --- the term computed in~\cite{Petcov:1976ff} --- while $f_1 \sim \log x_i$ is not, Ref.~\cite{Pham:1998fq} concludes that $(q^2/M_W^2) \log x_i $ must dominate the branching ratio in the $x_i \to 0$ limit.

This conclusion is flawed. First of all, in~\cite{Pham:1998fq} the loop integrals written in terms of Feynman parameters are computed via a simple Taylor expansion of the denominator appearing inside. Such approximation is not legitimate for arbitrary values of the Feynman parameters, so it does not lead to an expansion of the integral itself.
A proper asymptotic expansion of a Feynman integral can be obtained, for instance, via the expansion-by-regions method~\cite{Beneke:1997zp,Smirnov:2002pj}, in which one divides the whole integration domain into various regions and then performs different Taylor expansions in each region. Only the sum of all regions' contributions eventually yields the desired asymptotic series.
In addition to that, even if such $q^2/M_W^2$ expansion were performed correctly, the calculation presented in~\cite{Pham:1998fq} implicitly assumes also the hierarchy $q^2 \ll m_{\nu i}^2$. Therefore, the series expansion $f_0(x_i) + (q^2/M_W^2) f_1(x_i) + \dots$ does not reproduce the correct $x_i \to 0$ limit at fixed values of $q^2$, since this limit lies beyond the validity range of $q^2 \ll m_{\nu i}^2$. 

Recently, Ref.~\cite{Hernandez-Tome:2018fbq} presented a calculation of $L\to \ell \ell' \ell'$ in the PL, in which the one-loop diagrams are numerically evaluated with full dependence on $\mathcal{P}$. At variance with~\cite{Pham:1998fq}, they found branching ratios compatible with those in the ZML or smaller. 
However, the authors of Ref.~\cite{Hernandez-Tome:2018fbq} neglect the contribution from $\gamma$-penguins (as in Fig.~\ref{fig:Apen}) and therefore their results are gauge dependent. 
Indeed, in processes with flavour changing neutral currents the gauge cancels entirely only in the sum of boxes, $Z$- and $\gamma$-penguins~\cite{Inami:1980fz,Buchalla:1990qz}.
Even if Ref.~\cite{Petcov:1976ff} retained only the logarithmic enhanced term $x_i \log x_i$ arising only from the $Z$ penguins and the boxes, the omission of $\gamma$-penguins is not legitimate anymore as soon as one departs from this approximation. 
So we are still left with the doubt if the branching ratios in~\cite{Hernandez-Tome:2018fbq} are smaller as a consequence of calculating in the Feynman gauge or if there is a deeper physical meaning.

In this letter we present the decay widths of $L \to \ell \ell' \ell'$ in the PL, fully analytic in $M_W$, $m_\nu$ and  external momenta and masses.
We compute them by making a systematic asymptotic expansion in $\mathcal{P}/M_W$ and $m_\nu/\mathcal{P}$ of all Feynman diagrams by means of the expansion by regions.
We will show that the neutrino mass dependence $|\sum_i x_i \log x_i|^2$ in the ZML is replaced in the  PL by a much smaller enhancement $|\sum_i x_i \log(m_L^2/M_W^2)|^2$. We will give an explanation of this exchange of mass scales in the logarithm by analysing the effective operators mediating the decay once the $Z$ and the $W$ bosons are integrated out.

\section{Details of the calculation}
\begin{table*}[t]
\begin{ruledtabular}
  \begin{tabular}{c|ccc|ccc}
   & \multicolumn{3}{c}{Branching ratio (NO)}
   & \multicolumn{3}{c}{Branching ratio (IO)}\\
   & ZML & PL & ZML/PL & ZML & PL & ZML/PL\\
   \hline
   $\mu \to eee$ &  
   $4.1 \times 10^{-54}$ &  $2.9 \times 10^{-55}$ & 14 &
   $6.1 \times 10^{-54}$ &  $4.6 \times 10^{-55}$ & 14 \\
   $\tau \to \mu \mu \mu$ & 
   $2.0 \times 10^{-53}$ & $5.8 \times 10^{-55}$ & 34 &
   $2.0 \times 10^{-53}$ & $5.8 \times 10^{-55}$ & 34\\
   $\tau \to \mu e e $ & 
   $1.3 \times 10^{-53}$ & $3.8 \times 10^{-55}$ & 35 &
   $1.3 \times 10^{-53}$ & $ 3.8 \times 10^{-55}$ & 35 \\
   $\tau \to eee$ & 
   $1.1 \times 10^{-54}$ & $3.3 \times 10^{-56}$ & 34 &
   $6.1 \times 10^{-55}$ & $1.9 \times 10^{-56}$ & 32 \\
   $\tau \to e\mu\mu$ & 
   $ 7.6 \times 10^{-55}$ & $2.1 \times 10^{-56}$ & 36 &
   $4.1 \times 10^{-55}$ & $ 1.2 \times 10^{-56}$ & 34 \\
\end{tabular}
\end{ruledtabular}
\caption{Branching ratio for the CLFV decays $L \to \ell \ell'\ell'$ in the ZML and the PL for normal  ordering (NO) and inverted ordering (IO) of neutrino masses. 
The ratio between the two is also reported.
In the ZML we assume $m_1 = 0$ (NO) or $m_3=0$ (IO).}
  \label{tab:BRfinal}
\end{table*}
Let us consider the SM extended with neutrino masses of either a Dirac or Majorana nature. The flavour eigenstates of the left-handed neutrino fields $\nu_{\ell L}$ entering in the weak interactions become linear combinations of the three mass eigenstates $\nu_i$ with masses $m_{\nu i}$:
\begin{equation}
  \nu_{\ell L} = \sum_{i=1}^3 U_{\ell i} \nu_{i L}, 
  \quad \ell = e,\mu,\tau,
\end{equation}
where $\nu_{i L}$ is the left-handed component of $\nu_i$ and $U$ is the PMNS matrix. 
The decay of a heavy lepton $L=\mu,\tau$ into three lighter charged leptons $\ell,\ell'=\mu,e$,
\begin{equation}
  L^\pm \to \ell^\pm \ell'^+ \ell'^-,
  \label{eqn:L2lll}
\end{equation}
with masses $m_L$, $m_\ell$ and $m_{\ell'}$, respectively,  proceeds then via three classes of one-loop diagrams shown in Fig.~\ref{fig:fd}: the boxes, the $Z$ and $\gamma$ penguins. We neglect diagrams with the exchange of a Higgs boson, as they are further suppressed by two extra powers of $1/M_W^2$ due the Yukawa interaction. In the case of $L \to \ell \ell'\ell'$ there are additional box diagrams with the incoming $L$ line connected to the outgoing $\ell'$ line, and the outgoing $\ell$ line to the incoming $\ell'^{-}$ line. They can be neglected since they have two instances of CLFV and thus two of the GIM-breaking factors. 
The partial width given in~\cite{Petcov:1976ff} was obtained in the ZML. In this approximation, the one-loop integrals depend only on the ratio $x_i$. To leading order in $x_i$, the amplitudes of the three classes of diagrams are:
\begin{align}
      i\mathcal{A}_\gamma  &=
  -\frac{\alpha G_F}{ \sqrt{2} \pi} 
  (\bar{\ell} \gamma^\mu P_L L)
  (\bar{\ell'} \gamma_{\mu} \ell')
  \sum_i
  U_{\ell i} U^*_{L i} \, 
  x_i ,\\
  i\mathcal{A}_Z & =
  \frac{\alpha G_F}{\sqrt{2} \pi}
  \sum_i
  U_{\ell i} U^*_{L i} \, 
  x_i (3 + \log x_i) \notag \\
   \times 
  \Bigg[
  (\bar{\ell} &\gamma^\mu P_L L)
  (\bar{\ell'} \gamma_{\mu} \ell')
  -\frac{1}{2\sin^2 \theta_W}
  (\bar{\ell} \gamma^\mu P_L L)
  (\bar{\ell'} \gamma_{ \mu} P_L\ell') 
  \Bigg],
 \\
  i\mathcal{A}_\mathrm{Box} & =
  \frac{\alpha G_F}{ 2 \sqrt{2} \pi \sin^2 \theta_W} 
  (\bar{\ell} \gamma^\mu P_L L)
  (\bar{\ell'} \gamma_{\mu } P_L \ell') \notag \\
  & \quad \times \sum_i
  U_{\ell i} U^*_{L i} \, 
  x_i (1+\log x_i), 
  \label{eqn:ampZML}
\end{align}
where $G_F$ and $\alpha$ are the Fermi and fine structure constants, respectively, and $\sin^2 \theta_W$ is the sine of the Weinberg angle. Retaining only the terms enhanced by $\log x_i \sim 50$, which appear in the boxes and $Z$ penguins, one obtains the prediction for the rate in the ZML~\cite{Petcov:1976ff}:
\begin{equation}
  \frac{\Gamma(L \to \ell \ell \ell)}{\Gamma_0} =
  \frac{3 \alpha^2}{16\pi^2}
  \left\vert 
    \sum_{i=2}^3
    U_{\ell i} U^*_{L i}
    \frac{\Delta m_{i1}^2}{M_W^2}
    \log \frac{\Delta m_{i1}^2}{M_W^2}
  \right\vert^2 ,\label{eqn:GammaZML}
\end{equation}
and $
  \Gamma(L \to \ell \ell' \ell', \ell \neq \ell') = 
  \frac{2}{3} \Gamma(L \to \ell \ell \ell)
$,
where $\Gamma_0 = G_F^2 m_L^5/(192 \pi^3)$. Eq.~\eqref{eqn:GammaZML} is obtained by taking the limit $m_1 \to 0$ and assumes normal neutrino mass hierarchy, i.e.\ $m_1 < m_2 < m_3$.
For inverted mass hierarchy, the subscript `1' must be substituted with `3' and $i=1,2$. Eq.~\eqref{eqn:GammaZML} also neglects subleading $m_{\ell,\ell'}/m_L$ corrections from phase space integration. 
The values of the branching ratios in the ZML with normal and inverted mass hierarchy are reported in Tab.~\ref{tab:BRfinal}. Current PDG values are employed for the lepton masses, neutrino mass splittings and neutrino mixing angles~\cite{Tanabashi:2018oca}. 

Let us now describe our calculation performed in the PL. We generated the complete set of diagrams in the Feynman gauge, and their relative counter-terms, using \textsc{FeynArts}~\cite{Hahn:2000kx} with a modified version of the SM file to account for Dirac neutrino masses and lepton flavour mixing. 
The amplitudes were reduced to one-loop tensor integrals using \textsc{Form}~\cite{Kuipers:2012rf}, via the \textsc{FormCalc} package~\cite{Hahn:1998yk}, keeping the complete dependence on $M_W$, $\mathcal{P}$ and $m_{\nu i}$. The setup was independently checked by a second implementation based on \textsc{FeynCalc}~\cite{Shtabovenko:2016sxi}. 

Nowadays, lengthy expressions for the tensor integrals could be obtained in principle in an analytical form with full dependence on $m_{\nu i}, M_W, m_L, m_{\ell,\ell'}$ and the invariants $s_{ij} = (p_i+p_j)^2$, with $p_{1-3}$ the momenta of the three outgoing leptons, however their use is prohibitively cumbersome. It is therefore more helpful to compute them as series in the small parameters $\mathcal{P}^2/M_W^2$ and $m_{\nu i}^2/\mathcal{P}^2$. 
To this end, we employed the method of expansion by regions (for an introduction see e.g.~\cite{Smirnov:2002pj}). 
For all one-loop diagrams, we divided the integration domain into different regions and, for each region, we performed a Taylor expansion with respect to the parameters that are considered small there. 
Afterwards, by integrating every expanded integrand over the whole domain, and by summing the contributions from all the regions, we obtained the desired asymptotic expansion of the original one-loop diagram.
The advantage of this method, compared for instance to an expansion of the full result \emph{after} integration, is that the integrals arising in each region can be handled much more easily than the initial one, as typically they depend on just one or two mass scales.

We performed first an expansion assuming $\mathcal{P} \sim m_{\nu _i} \ll M_W$,  without distinguishing at this point the two scales $m_{\nu i}$ and $\mathcal{P}$.
In a second step, the integrals arising from the first stage are further expanded in the limit $m_{\nu i} \ll \mathcal{P}$. 
The total amplitude is then obtained by retaining from this expansion only the leading dependence on $m_{\nu i}$, while higher order terms further suppressed by $\mathcal{P}^2/M_W^2$ or $m_{\nu i }^2/\mathcal{P}^2$, or terms independent  on $m_{\nu i}$, are discarded. 
We performed several numerical checks at different stages of the calculation as a sanity check. To this end we took advantage of \textsc{Mathematica}’s arbitrary-precision numbers and \textsc{Package-X}'s analytic expressions of one-loop integrals~\cite{Patel:2015tea}, available in any kinematic configuration.
We verified that our approximated expressions for the tensor integrals became increasingly accurate both by including higher order terms in the expansion as well as by taking the limit $M_W \to \infty$ and $m_{\nu i} \to 0$, at fixed values of $\mathcal{P}$. 

\section{Results}
The partial widths are given by integrating the squared amplitude over the three-particle phase space of $L\to \ell \ell'\ell'$. 
These massive phase space integrals depend on two variables $s_{ij}$, plus two or three masses of the external particles. 
By employing the expansion by regions one more time, we computed the phase space integrals as series in $m_{\ell^{(')}}/m_L$, retaining only the leading terms in the final expressions for the rate. 
We obtain for normal neutrino mass hierarchy: 
\begin{multline}
  \frac{\Gamma (L \to \ell \ell\ell)}{\Gamma_0} =  
   \frac{3 \alpha^2}{16 \pi^2} 
   \left\vert
   \sum_{i=2}^3 U_{\ell i} U_{L i}^* \frac{\Delta m_{i1}^2}{M_W^2} 
   \right\vert^2   \\
   \times \Bigg[ 
    \log^2 x_L  
    +2 \log x_L   
    -\frac{1}{6}\log x_\ell  
    +\frac{19}{18}
    +\frac{17}{18}\pi^2 \\ \qquad \qquad
    -\frac{1}{\sin^2 \theta_W} 
    \left(\log x_L +  \frac{11}{12}  \right)
    +\frac{3}{8 \sin^4 \theta_W} 
  \Bigg], 
  \label{eqn:GammaPL}
\end{multline}
where $x_L = m_L^2/M_W^2$ and $x_\ell = m_\ell^2/M_W^2$.
For $L \to \ell \ell'\ell'$ ($\ell \neq \ell'$) we have:
\begin{multline}
  \frac{\Gamma (L \to \ell \ell'\ell')}{\Gamma_0} =  
   \frac{3 \alpha^2}{16 \pi^2} 
   \left\vert
   \sum_{i=2}^3 U_{\ell i} U_{L i}^* \frac{\Delta m_{i1}^2}{M_W^2} 
   \right\vert^2 \\
    \times \Bigg[ 
     \frac{2}{3} \log^2 x_L  
    +\frac{25}{18} \log x_L   
    -\frac{1}{6} \log x_{\ell'}  
    +\frac{55}{108}+\frac{2}{3}\pi^2 \\
    -\frac{1}{\sin^2 \theta_W} 
    \left(
    \frac{\log x_L}{2} + \frac{11}{24}+\frac{\pi^2}{18} 
    \right)\\
    +\frac{1}{\sin^4 \theta_W}
    \left( 
      \frac{3}{16}+\frac{\pi^2}{36}
    \right)
  \Bigg].
  \label{eqn:GammaPL2}
\end{multline}
Eqs.~\eqref{eqn:GammaPL} and \eqref{eqn:GammaPL2} depend only on the neutrino mass splittings $\Delta m_{ij}^2$ and not on the value of the lightest neutrino's mass. The expressions with inverted neutrino hierarchy are obtained similarly as for the ZML.
Higher order terms not included in Eqs.~\eqref{eqn:GammaPL} and \eqref{eqn:GammaPL2} are suppressed by $m_L^2/M_W^2$ or $m_{\nu i}^2/m_L^2$. These corrections would arise by further expanding the squared amplitude. In addition, there are also subleading $m_\ell/m_L$ terms from the phase space integration.

At variance with the results presented in~\cite{Pham:1998fq}, Eqs.~\eqref{eqn:GammaPL} and \eqref{eqn:GammaPL2}  are power suppressed  by $\vert \sum_i U_{\ell i} U_{L i}^* x_i\vert^2$ and yield values for the branching ratios of the order of $10^{-55}$, see Tab.~\ref{tab:BRfinal}. 
Moreover, compared to Eq.~\eqref{eqn:GammaZML} in the ZML, they do not have a logarithmic enhancement $\log^2 x_i \sim 2500$. On the contrary in its place we get only $\log^2 x_\mu \sim 176$ or $\log^2 x_\tau \sim 58$, which are of comparable size with respect to other terms appearing in Eqs.~\eqref{eqn:GammaPL} and ~\eqref{eqn:GammaPL2}. For this reason, the branching ratios in the PL turn out to be about one order of magnitude smaller than those in the ZML. 

Note also that the presence of the singular terms $\log x_{\ell}$ or $\log x_{\ell'}$ is not in contradiction with the Kinoshita-Lee-Nauenberg theorem~\cite{Kinoshita:1962ur,Lee:1964is} and the cancellation of mass singularities for inclusive observables. In fact Eqs.~\eqref{eqn:GammaPL} and ~\eqref{eqn:GammaPL2} are valid strictly in the PL, i.e.\ when $m_{\nu i} \ll m_{\ell^{(')}}$. The limiting case of vanishing charged-lepton masses and non-zero neutrino masses violates the assumptions of our derivation and therefore is not a meaningful limit of our expressions.

In the course of this letter, we have treated the neutrinos' masses as being Dirac in nature. This is appropriate towards the goal of evaluating the claim of \cite{Pham:1998fq}. The further evaluation of the case of Majorana masses requires additional particle content and implementation of a mechanism such as the well known seesaw mechanism~\cite{Minkowski:1977sc, GellMann:1980vs, Yanagida:1979as, Mohapatra:1979ia}. As such, this evaluation depends on the New Physics model being examined and lies beyond the scope of this SM calculation. 

\begin{figure}[thb]
    \centering
    \subfloat[\label{fig:fish}]{\includegraphics[height=0.25\columnwidth]{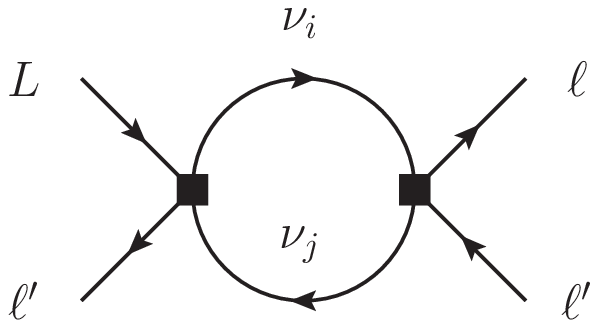}}\quad
    \subfloat[\label{gif:4f}]{\includegraphics[height=0.22\columnwidth]{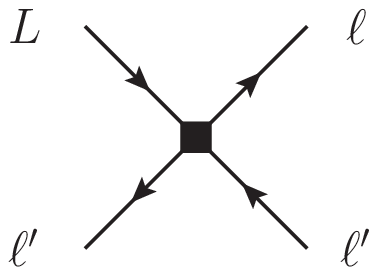}}
    \caption{Example of two diagrams mediating $L \to \ell \ell' \ell'$ in a low energy effective field theory description.}
    \label{fig:eft}
\end{figure}
Finally, we can understand the mechanism that converts the $ x_i \log x_i$ in the ZML into a $x_i \log x_L$ in the PL by looking at the underlying effective theory arising after integrating out the $Z$ and the $W$ bosons. For simplicity, let us concentrate only on the operators associated with such logarithmic enhancement in the box~\ref{fig:box}. 
Both in the ZML and in the PL, we can shrink the two $W$ propagators to a point-like interaction and match the amplitude onto the the following dimension-six and dimension-eight operators:
\begin{align}
    O_6^{L\ell'} &=
    (\bar{\nu}_i \gamma^\mu P_L L) \, (\bar{\ell}' \gamma_{\mu} P_ L \nu_j) \, , \notag \\
    O_6^{\ell\ell'} &=
    (\bar{\ell} \gamma^\mu P_L \nu_i) \, (\bar{\nu}_j \gamma_{\mu} P_L \ell')\, , \notag \\
    O_8 &= 
    m_{\nu i}^2 \, (\bar{\ell} \gamma^\mu P_L L) \, (\bar{\ell'} \gamma_\mu P_L \ell')\, .
    \label{eqn:opEFT}
\end{align}
The first two operators correspond to the usual Fermi interaction mediating $\mu$ and $\tau$ leptonic decays. They contribute to $L \to \ell \ell'\ell'$ via the one-loop diagram in Fig.~\ref{fig:fish}. 
The third operator in~\eqref{eqn:opEFT} is necessary to renormalize the effective theory, i.e.\ to cancel the UV divergence from the diagram~\ref{fig:fish}.
These operators' Wilson coefficients are:
\begin{gather}
    C_6^{L\ell'} = \frac{4 G_F}{\sqrt{2}} U_{\ell' j} U_{L i}^* \, ,  \qquad
    C_6^{\ell\ell'} = \frac{4 G_F}{\sqrt{2}} U_{\ell' j}^* U_{\ell i}\, , \notag \\
    C_8 (\mu) =  \frac{G_F^2}{2 \pi^2 } \log \left( \frac{M_W^2}{\mu^2}\right)
    U_{\ell i} U_{L i}^* \, .
\end{gather}
We can imagine performing the matching between the SM and the effective theory at a scale $\mu=M_W$, and evolving the coefficients to a lower scale via the renormalization group.
The coefficient $C_8$ explicitly depends on the renormalization scale $\mu$ and this dependence reveals the difference between the PL and the ZML. 
In the ZML, the evolution of $C_8$ can proceed down to a scale $\mu \sim m_{\nu i}$, at which point we can integrate out the neutrinos and remove the operators $O_6^{L\ell'}$ and $O_6^{\ell \ell'}$ which contain the neutrino field. We are then left with an effective theory with only $O_8$, whose Wilson coefficient is frozen at $C_8(m_{\nu i})$, i.e.\ it contains a $\log(M_W^2/m_{\nu i}^2)$ (compare with Eq.~\eqref{eqn:ampZML}). 
On the contrary, in the PL, $C_8$ can run only until the scale $\mu \sim m_L$ is reached.  
In this case, all operators in~\eqref{eqn:opEFT} are still active at the scale $m_L$, however $C_8$ produces only a milder $\log(M^2_W/m_L^2)$ enhancement.
Similar considerations can be applied as well to the $Z$ penguin~\ref{fig:Zpen}.
Therefore, the ZML overestimates the values for the branching ratios as it allows an unphysical evolution of these operators between $M_W$ and $m_{\nu i}$, while in reality the running stops at the physical intermediate scale $m_L$ where the process happens.

\section{Conclusions}
Several experimental collaborations reported that the branching fractions of $L\to \ell \ell'\ell'$ can be as large as $10^{-14}$, following the observation in~\cite{Pham:1998fq} that the GIM cancellation for these decays is not so severe and takes the form of $|\sum_i U_{\ell i} U^*_{Li} \log x_i|^2$. In this letter we showed that this conclusion is wrong.

We calculated and presented for the first time the branching ratios in the PL by performing a series expansion of all one-loop diagrams in the small parameters $\mathcal{P}/M_W$ and $m_{\nu i}/\mathcal{P}$. 
Our fully analytic expressions prove that the GIM suppression in these decays is power-like $|\sum_i U_{\ell i} U^*_{Li} x_i |^2$, similar to that found previously in~\cite{Petcov:1976ff} for the ZML, so that the claim from~\cite{Pham:1998fq} must be rejected. 
We predicted the branching ratios in the Standard Model including neutrino masses to be of the order of $10^{-55}$, even smaller than those obtained in the ZML, as the latter prediction contains an unphysical logarithmic enhancement.

In the end, we remark that since the GIM suppression is solely governed by the underlying effective description of the process, i.e.\ the hierarchy of the internal mass scales and the convergence properties of Feynman integrals, external momentum effects could not have affected the GIM cancellation to such a large extent, as claimed in Ref.~\cite{Pham:1998fq}, compared to the finding in the ZML.

\bigskip
\begin{acknowledgments}
 {\em Acknowledgments} 
 We are grateful to S. Banerjee for drawing our attention to the interest of this calculation. 
 We thank G. Hern\'{a}ndez-Tom\'{e}, G.\ L\'{o}pez Castro and P.\ Roig for useful correspondence and clarifications about their work.
 We wish to thank also A.\ Crivellin, A.\ Friedland, T.\ Huber, L.\ Lavoura, J.\ Piclum and A.\ Signer for useful discussions and correspondence.  
 P.B.\ thanks also the Theoretical Particle Physics Group at the University of Siegen for the hospitality during his stay in June 2019. 
 M.F.\  is  supported  by  the  Deutsche  Forschungsgemeinschaft (DFG, German Research Foundation) under grant 396021762 - TRR 257 ``Particle Physics Phenomenology after the Higgs Discovery''. The work of 
P.B.\ and E.P.\ is supported in part by the U.S. Department of Energy (contract DE-AC05-06OR23177) and National
Science Foundation (PHY-1714253). 
\end{acknowledgments}

\bibliographystyle{apsrev4-1}
\bibliography{BIB}

\begin{thebibliography}{36}%
\makeatletter
\providecommand \@ifxundefined [1]{%
 \@ifx{#1\undefined}
}%
\providecommand \@ifnum [1]{%
 \ifnum #1\expandafter \@firstoftwo
 \else \expandafter \@secondoftwo
 \fi
}%
\providecommand \@ifx [1]{%
 \ifx #1\expandafter \@firstoftwo
 \else \expandafter \@secondoftwo
 \fi
}%
\providecommand \natexlab [1]{#1}%
\providecommand \enquote  [1]{``#1''}%
\providecommand \bibnamefont  [1]{#1}%
\providecommand \bibfnamefont [1]{#1}%
\providecommand \citenamefont [1]{#1}%
\providecommand \href@noop [0]{\@secondoftwo}%
\providecommand \href [0]{\begingroup \@sanitize@url \@href}%
\providecommand \@href[1]{\@@startlink{#1}\@@href}%
\providecommand \@@href[1]{\endgroup#1\@@endlink}%
\providecommand \@sanitize@url [0]{\catcode `\\12\catcode `\$12\catcode
  `\&12\catcode `\#12\catcode `\^12\catcode `\_12\catcode `\%12\relax}%
\providecommand \@@startlink[1]{}%
\providecommand \@@endlink[0]{}%
\providecommand \url  [0]{\begingroup\@sanitize@url \@url }%
\providecommand \@url [1]{\endgroup\@href {#1}{\urlprefix }}%
\providecommand \urlprefix  [0]{URL }%
\providecommand \Eprint [0]{\href }%
\providecommand \doibase [0]{http://dx.doi.org/}%
\providecommand \selectlanguage [0]{\@gobble}%
\providecommand \bibinfo  [0]{\@secondoftwo}%
\providecommand \bibfield  [0]{\@secondoftwo}%
\providecommand \translation [1]{[#1]}%
\providecommand \BibitemOpen [0]{}%
\providecommand \bibitemStop [0]{}%
\providecommand \bibitemNoStop [0]{.\EOS\space}%
\providecommand \EOS [0]{\spacefactor3000\relax}%
\providecommand \BibitemShut  [1]{\csname bibitem#1\endcsname}%
\let\auto@bib@innerbib\@empty
\bibitem [{\citenamefont {Petcov}(1977)}]{Petcov:1976ff}%
  \BibitemOpen
  \bibfield  {author} {\bibinfo {author} {\bibfnamefont {S.~T.}\ \bibnamefont
  {Petcov}},\ }\href@noop {} {\bibfield  {journal} {\bibinfo  {journal} {Sov.
  J. Nucl. Phys.}\ }\textbf {\bibinfo {volume} {25}},\ \bibinfo {pages} {340}
  (\bibinfo {year} {1977})},\ \bibinfo {note} {[Erratum: Yad.
  Fiz.25,1336(1977)]}\BibitemShut {NoStop}%
\bibitem [{\citenamefont {Pham}(1999)}]{Pham:1998fq}%
  \BibitemOpen
  \bibfield  {author} {\bibinfo {author} {\bibfnamefont {X.-Y.}\ \bibnamefont
  {Pham}},\ }\href {\doibase 10.1007/s100529901088} {\bibfield  {journal}
  {\bibinfo  {journal} {Eur. Phys. J.}\ }\textbf {\bibinfo {volume} {C8}},\
  \bibinfo {pages} {513} (\bibinfo {year} {1999})},\ \Eprint
  {http://arxiv.org/abs/hep-ph/9810484} {arXiv:hep-ph/9810484 [hep-ph]}
  \BibitemShut {NoStop}%
\bibitem [{CMS(2019)}]{CMS:2019sxo}%
  \BibitemOpen
  \href@noop {} {\emph {\bibinfo {title} {{Search for $\tau \to 3\mu$ decays
  using $\tau$ leptons produced in D and B meson decays}}}},\ \bibinfo {type}
  {Tech. Rep.}\ \bibinfo {number} {CMS-PAS-BPH-17-004}\ (\bibinfo
  {institution} {CERN},\ \bibinfo {address} {Geneva},\ \bibinfo {year}
  {2019})\BibitemShut {NoStop}%
\bibitem [{\citenamefont {Aad}\ \emph {et~al.}(2016)\citenamefont {Aad} \emph
  {et~al.}}]{Aad:2016wce}%
  \BibitemOpen
  \bibfield  {author} {\bibinfo {author} {\bibfnamefont {G.}~\bibnamefont
  {Aad}} \emph {et~al.} (\bibinfo {collaboration} {ATLAS}),\ }\href {\doibase
  10.1140/epjc/s10052-016-4041-9} {\bibfield  {journal} {\bibinfo  {journal}
  {Eur. Phys. J.}\ }\textbf {\bibinfo {volume} {C76}},\ \bibinfo {pages} {232}
  (\bibinfo {year} {2016})},\ \Eprint {http://arxiv.org/abs/1601.03567}
  {arXiv:1601.03567 [hep-ex]} \BibitemShut {NoStop}%
\bibitem [{\citenamefont {Aaij}\ \emph {et~al.}(2015)\citenamefont {Aaij} \emph
  {et~al.}}]{Aaij:2014azz}%
  \BibitemOpen
  \bibfield  {author} {\bibinfo {author} {\bibfnamefont {R.}~\bibnamefont
  {Aaij}} \emph {et~al.} (\bibinfo {collaboration} {LHCb}),\ }\href {\doibase
  10.1007/JHEP02(2015)121} {\bibfield  {journal} {\bibinfo  {journal} {JHEP}\
  }\textbf {\bibinfo {volume} {02}},\ \bibinfo {pages} {121} (\bibinfo {year}
  {2015})},\ \Eprint {http://arxiv.org/abs/1409.8548} {arXiv:1409.8548
  [hep-ex]} \BibitemShut {NoStop}%
\bibitem [{\citenamefont {Aubert}\ \emph {et~al.}(2004)\citenamefont {Aubert}
  \emph {et~al.}}]{Aubert:2003pc}%
  \BibitemOpen
  \bibfield  {author} {\bibinfo {author} {\bibfnamefont {B.}~\bibnamefont
  {Aubert}} \emph {et~al.} (\bibinfo {collaboration} {BaBar}),\ }\href
  {\doibase 10.1103/PhysRevLett.92.121801} {\bibfield  {journal} {\bibinfo
  {journal} {Phys. Rev. Lett.}\ }\textbf {\bibinfo {volume} {92}},\ \bibinfo
  {pages} {121801} (\bibinfo {year} {2004})},\ \Eprint
  {http://arxiv.org/abs/hep-ex/0312027} {arXiv:hep-ex/0312027 [hep-ex]}
  \BibitemShut {NoStop}%
\bibitem [{\citenamefont {Aubert}\ \emph {et~al.}(2007)\citenamefont {Aubert}
  \emph {et~al.}}]{Aubert:2007pw}%
  \BibitemOpen
  \bibfield  {author} {\bibinfo {author} {\bibfnamefont {B.}~\bibnamefont
  {Aubert}} \emph {et~al.} (\bibinfo {collaboration} {BaBar}),\ }\href
  {\doibase 10.1103/PhysRevLett.99.251803} {\bibfield  {journal} {\bibinfo
  {journal} {Phys. Rev. Lett.}\ }\textbf {\bibinfo {volume} {99}},\ \bibinfo
  {pages} {251803} (\bibinfo {year} {2007})},\ \Eprint
  {http://arxiv.org/abs/0708.3650} {arXiv:0708.3650 [hep-ex]} \BibitemShut
  {NoStop}%
\bibitem [{\citenamefont {Aubert}\ \emph {et~al.}(2008)\citenamefont {Aubert}
  \emph {et~al.}}]{Aubert:2007kx}%
  \BibitemOpen
  \bibfield  {author} {\bibinfo {author} {\bibfnamefont {B.}~\bibnamefont
  {Aubert}} \emph {et~al.} (\bibinfo {collaboration} {BaBar}),\ }\href
  {\doibase 10.1103/PhysRevLett.100.071802} {\bibfield  {journal} {\bibinfo
  {journal} {Phys. Rev. Lett.}\ }\textbf {\bibinfo {volume} {100}},\ \bibinfo
  {pages} {071802} (\bibinfo {year} {2008})},\ \Eprint
  {http://arxiv.org/abs/0711.0980} {arXiv:0711.0980 [hep-ex]} \BibitemShut
  {NoStop}%
\bibitem [{\citenamefont {Lees}\ \emph {et~al.}(2010)\citenamefont {Lees} \emph
  {et~al.}}]{Lees:2010ez}%
  \BibitemOpen
  \bibfield  {author} {\bibinfo {author} {\bibfnamefont {J.~P.}\ \bibnamefont
  {Lees}} \emph {et~al.} (\bibinfo {collaboration} {BaBar}),\ }\href {\doibase
  10.1103/PhysRevD.81.111101} {\bibfield  {journal} {\bibinfo  {journal} {Phys.
  Rev.}\ }\textbf {\bibinfo {volume} {D81}},\ \bibinfo {pages} {111101}
  (\bibinfo {year} {2010})},\ \Eprint {http://arxiv.org/abs/1002.4550}
  {arXiv:1002.4550 [hep-ex]} \BibitemShut {NoStop}%
\bibitem [{\citenamefont {Hayasaka}\ \emph {et~al.}(2010)\citenamefont
  {Hayasaka} \emph {et~al.}}]{Hayasaka:2010np}%
  \BibitemOpen
  \bibfield  {author} {\bibinfo {author} {\bibfnamefont {K.}~\bibnamefont
  {Hayasaka}} \emph {et~al.},\ }\href {\doibase 10.1016/j.physletb.2010.03.037}
  {\bibfield  {journal} {\bibinfo  {journal} {Phys. Lett.}\ }\textbf {\bibinfo
  {volume} {B687}},\ \bibinfo {pages} {139} (\bibinfo {year} {2010})},\ \Eprint
  {http://arxiv.org/abs/1001.3221} {arXiv:1001.3221 [hep-ex]} \BibitemShut
  {NoStop}%
\bibitem [{\citenamefont {Glashow}\ \emph {et~al.}(1970)\citenamefont
  {Glashow}, \citenamefont {Iliopoulos},\ and\ \citenamefont
  {Maiani}}]{Glashow:1970gm}%
  \BibitemOpen
  \bibfield  {author} {\bibinfo {author} {\bibfnamefont {S.~L.}\ \bibnamefont
  {Glashow}}, \bibinfo {author} {\bibfnamefont {J.}~\bibnamefont {Iliopoulos}},
  \ and\ \bibinfo {author} {\bibfnamefont {L.}~\bibnamefont {Maiani}},\
  }\bibfield  {booktitle} {\emph {\bibinfo {booktitle} {{Meeting of the Italian
  School of Physics and Weak Interactions Bologna, Italy, April 26-28,
  1984}}},\ }\href {\doibase 10.1103/PhysRevD.2.1285} {\bibfield  {journal}
  {\bibinfo  {journal} {Phys. Rev.}\ }\textbf {\bibinfo {volume} {D2}},\
  \bibinfo {pages} {1285} (\bibinfo {year} {1970})}\BibitemShut {NoStop}%
\bibitem [{\citenamefont {Pontecorvo}(1968)}]{Pontecorvo:1967fh}%
  \BibitemOpen
  \bibfield  {author} {\bibinfo {author} {\bibfnamefont {B.}~\bibnamefont
  {Pontecorvo}},\ }\href@noop {} {\bibfield  {journal} {\bibinfo  {journal}
  {Sov. Phys. JETP}\ }\textbf {\bibinfo {volume} {26}},\ \bibinfo {pages} {984}
  (\bibinfo {year} {1968})},\ \bibinfo {note} {[Zh. Eksp. Teor.
  Fiz.53,1717(1967)]}\BibitemShut {NoStop}%
\bibitem [{\citenamefont {Maki}\ \emph {et~al.}(1962)\citenamefont {Maki},
  \citenamefont {Nakagawa},\ and\ \citenamefont {Sakata}}]{Maki:1962mu}%
  \BibitemOpen
  \bibfield  {author} {\bibinfo {author} {\bibfnamefont {Z.}~\bibnamefont
  {Maki}}, \bibinfo {author} {\bibfnamefont {M.}~\bibnamefont {Nakagawa}}, \
  and\ \bibinfo {author} {\bibfnamefont {S.}~\bibnamefont {Sakata}},\ }\href
  {\doibase 10.1143/PTP.28.870} {\bibfield  {journal} {\bibinfo  {journal}
  {Prog. Theor. Phys.}\ }\textbf {\bibinfo {volume} {28}},\ \bibinfo {pages}
  {870} (\bibinfo {year} {1962})}\BibitemShut {NoStop}%
\bibitem [{\citenamefont {Tanabashi}\ \emph {et~al.}(2018)\citenamefont
  {Tanabashi} \emph {et~al.}}]{Tanabashi:2018oca}%
  \BibitemOpen
  \bibfield  {author} {\bibinfo {author} {\bibfnamefont {M.}~\bibnamefont
  {Tanabashi}} \emph {et~al.} (\bibinfo {collaboration} {Particle Data
  Group}),\ }\href {\doibase 10.1103/PhysRevD.98.030001} {\bibfield  {journal}
  {\bibinfo  {journal} {Phys. Rev.}\ }\textbf {\bibinfo {volume} {D98}},\
  \bibinfo {pages} {030001} (\bibinfo {year} {2018})}\BibitemShut {NoStop}%
\bibitem [{\citenamefont {Blondel}\ \emph {et~al.}(2013)\citenamefont {Blondel}
  \emph {et~al.}}]{Blondel:2013ia}%
  \BibitemOpen
  \bibfield  {author} {\bibinfo {author} {\bibfnamefont {A.}~\bibnamefont
  {Blondel}} \emph {et~al.},\ }\href@noop {} {\  (\bibinfo {year} {2013})},\
  \Eprint {http://arxiv.org/abs/1301.6113} {arXiv:1301.6113 [physics.ins-det]}
  \BibitemShut {NoStop}%
\bibitem [{\citenamefont {Amhis}\ \emph {et~al.}(2019)\citenamefont {Amhis}
  \emph {et~al.}}]{Amhis:2019ckw}%
  \BibitemOpen
  \bibfield  {author} {\bibinfo {author} {\bibfnamefont {Y.~S.}\ \bibnamefont
  {Amhis}} \emph {et~al.} (\bibinfo {collaboration} {HFLAV}),\ }\href@noop {}
  {\  (\bibinfo {year} {2019})},\ \Eprint {http://arxiv.org/abs/1909.12524}
  {arXiv:1909.12524 [hep-ex]} \BibitemShut {NoStop}%
\bibitem [{\citenamefont {Altmannshofer}\ \emph {et~al.}(2018)\citenamefont
  {Altmannshofer} \emph {et~al.}}]{Kou:2018nap}%
  \BibitemOpen
  \bibfield  {author} {\bibinfo {author} {\bibfnamefont {W.}~\bibnamefont
  {Altmannshofer}} \emph {et~al.} (\bibinfo {collaboration} {Belle-II}),\
  }\href@noop {} {\  (\bibinfo {year} {2018})},\ \Eprint
  {http://arxiv.org/abs/1808.10567} {arXiv:1808.10567 [hep-ex]} \BibitemShut
  {NoStop}%
\bibitem [{\citenamefont {Cerri}\ \emph {et~al.}(2018)\citenamefont {Cerri}
  \emph {et~al.}}]{Cerri:2018ypt}%
  \BibitemOpen
  \bibfield  {author} {\bibinfo {author} {\bibfnamefont {A.}~\bibnamefont
  {Cerri}} \emph {et~al.},\ }\href@noop {} {\  (\bibinfo {year} {2018})},\
  \Eprint {http://arxiv.org/abs/1812.07638} {arXiv:1812.07638 [hep-ph]}
  \BibitemShut {NoStop}%
\bibitem [{\citenamefont {Fiorendi}(2018)}]{Fiorendi:2018qwm}%
  \BibitemOpen
  \bibfield  {author} {\bibinfo {author} {\bibfnamefont {S.}~\bibnamefont
  {Fiorendi}} (\bibinfo {collaboration} {CMS}),\ }\bibfield  {booktitle} {\emph
  {\bibinfo {booktitle} {{Proceedings, 17th International Conference on
  B-Physics at Frontier Machines (Beauty 2018): La Biodola, Elba island, Italy,
  May 6-11, 2018}}},\ }\href {\doibase 10.22323/1.326.0056} {\bibfield
  {journal} {\bibinfo  {journal} {PoS}\ }\textbf {\bibinfo {volume}
  {BEAUTY2018}},\ \bibinfo {pages} {056} (\bibinfo {year} {2018})}\BibitemShut
  {NoStop}%
\bibitem [{\citenamefont {Azzi}\ \emph {et~al.}(2019)\citenamefont {Azzi} \emph
  {et~al.}}]{Azzi:2019yne}%
  \BibitemOpen
  \bibfield  {author} {\bibinfo {author} {\bibfnamefont {P.}~\bibnamefont
  {Azzi}} \emph {et~al.} (\bibinfo {collaboration} {HL-LHC, HE-LHC Working
  Group}),\ }\href@noop {} {\  (\bibinfo {year} {2019})},\ \Eprint
  {http://arxiv.org/abs/1902.04070} {arXiv:1902.04070 [hep-ph]} \BibitemShut
  {NoStop}%
\bibitem [{\citenamefont {Beneke}\ and\ \citenamefont
  {Smirnov}(1998)}]{Beneke:1997zp}%
  \BibitemOpen
  \bibfield  {author} {\bibinfo {author} {\bibfnamefont {M.}~\bibnamefont
  {Beneke}}\ and\ \bibinfo {author} {\bibfnamefont {V.~A.}\ \bibnamefont
  {Smirnov}},\ }\href {\doibase 10.1016/S0550-3213(98)00138-2} {\bibfield
  {journal} {\bibinfo  {journal} {Nucl. Phys.}\ }\textbf {\bibinfo {volume}
  {B522}},\ \bibinfo {pages} {321} (\bibinfo {year} {1998})},\ \Eprint
  {http://arxiv.org/abs/hep-ph/9711391} {arXiv:hep-ph/9711391 [hep-ph]}
  \BibitemShut {NoStop}%
\bibitem [{\citenamefont {Smirnov}(2002)}]{Smirnov:2002pj}%
  \BibitemOpen
  \bibfield  {author} {\bibinfo {author} {\bibfnamefont {V.~A.}\ \bibnamefont
  {Smirnov}},\ }\bibfield  {booktitle} {\emph {\bibinfo {booktitle} {{Berlin,
  Germany: Springer (2002) 262 p, (Springer tracts in modern physics. 177)}}},\
  }\href@noop {} {\bibfield  {journal} {\bibinfo  {journal} {Springer Tracts
  Mod. Phys.}\ }\textbf {\bibinfo {volume} {177}},\ \bibinfo {pages} {1}
  (\bibinfo {year} {2002})}\BibitemShut {NoStop}%
\bibitem [{\citenamefont {Hernandez-Tome}\ \emph {et~al.}(2019)\citenamefont
  {Hernandez-Tome}, \citenamefont {Lopez~Castro},\ and\ \citenamefont
  {Roig}}]{Hernandez-Tome:2018fbq}%
  \BibitemOpen
  \bibfield  {author} {\bibinfo {author} {\bibfnamefont {G.}~\bibnamefont
  {Hernandez-Tome}}, \bibinfo {author} {\bibfnamefont {G.}~\bibnamefont
  {Lopez~Castro}}, \ and\ \bibinfo {author} {\bibfnamefont {P.}~\bibnamefont
  {Roig}},\ }\href {\doibase 10.1140/epjc/s10052-019-6563-4} {\bibfield
  {journal} {\bibinfo  {journal} {Eur. Phys. J.}\ }\textbf {\bibinfo {volume}
  {C79}},\ \bibinfo {pages} {84} (\bibinfo {year} {2019})},\ \Eprint
  {http://arxiv.org/abs/1807.06050} {arXiv:1807.06050 [hep-ph]} \BibitemShut
  {NoStop}%
\bibitem [{\citenamefont {Inami}\ and\ \citenamefont
  {Lim}(1981)}]{Inami:1980fz}%
  \BibitemOpen
  \bibfield  {author} {\bibinfo {author} {\bibfnamefont {T.}~\bibnamefont
  {Inami}}\ and\ \bibinfo {author} {\bibfnamefont {C.~S.}\ \bibnamefont
  {Lim}},\ }\href {\doibase 10.1143/PTP.65.297} {\bibfield  {journal} {\bibinfo
   {journal} {Prog. Theor. Phys.}\ }\textbf {\bibinfo {volume} {65}},\ \bibinfo
  {pages} {297} (\bibinfo {year} {1981})},\ \bibinfo {note} {[Erratum: Prog.
  Theor. Phys.65,1772(1981)]}\BibitemShut {NoStop}%
\bibitem [{\citenamefont {Buchalla}\ \emph {et~al.}(1991)\citenamefont
  {Buchalla}, \citenamefont {Buras},\ and\ \citenamefont
  {Harlander}}]{Buchalla:1990qz}%
  \BibitemOpen
  \bibfield  {author} {\bibinfo {author} {\bibfnamefont {G.}~\bibnamefont
  {Buchalla}}, \bibinfo {author} {\bibfnamefont {A.~J.}\ \bibnamefont {Buras}},
  \ and\ \bibinfo {author} {\bibfnamefont {M.~K.}\ \bibnamefont {Harlander}},\
  }\href {\doibase 10.1016/0550-3213(91)90186-2} {\bibfield  {journal}
  {\bibinfo  {journal} {Nucl. Phys.}\ }\textbf {\bibinfo {volume} {B349}},\
  \bibinfo {pages} {1} (\bibinfo {year} {1991})}\BibitemShut {NoStop}%
\bibitem [{\citenamefont {Hahn}(2001)}]{Hahn:2000kx}%
  \BibitemOpen
  \bibfield  {author} {\bibinfo {author} {\bibfnamefont {T.}~\bibnamefont
  {Hahn}},\ }\href {\doibase 10.1016/S0010-4655(01)00290-9} {\bibfield
  {journal} {\bibinfo  {journal} {Comput. Phys. Commun.}\ }\textbf {\bibinfo
  {volume} {140}},\ \bibinfo {pages} {418} (\bibinfo {year} {2001})},\ \Eprint
  {http://arxiv.org/abs/hep-ph/0012260} {arXiv:hep-ph/0012260 [hep-ph]}
  \BibitemShut {NoStop}%
\bibitem [{\citenamefont {Kuipers}\ \emph {et~al.}(2013)\citenamefont
  {Kuipers}, \citenamefont {Ueda}, \citenamefont {Vermaseren},\ and\
  \citenamefont {Vollinga}}]{Kuipers:2012rf}%
  \BibitemOpen
  \bibfield  {author} {\bibinfo {author} {\bibfnamefont {J.}~\bibnamefont
  {Kuipers}}, \bibinfo {author} {\bibfnamefont {T.}~\bibnamefont {Ueda}},
  \bibinfo {author} {\bibfnamefont {J.~A.~M.}\ \bibnamefont {Vermaseren}}, \
  and\ \bibinfo {author} {\bibfnamefont {J.}~\bibnamefont {Vollinga}},\ }\href
  {\doibase 10.1016/j.cpc.2012.12.028} {\bibfield  {journal} {\bibinfo
  {journal} {Comput. Phys. Commun.}\ }\textbf {\bibinfo {volume} {184}},\
  \bibinfo {pages} {1453} (\bibinfo {year} {2013})},\ \Eprint
  {http://arxiv.org/abs/1203.6543} {arXiv:1203.6543 [cs.SC]} \BibitemShut
  {NoStop}%
\bibitem [{\citenamefont {Hahn}\ and\ \citenamefont
  {Perez-Victoria}(1999)}]{Hahn:1998yk}%
  \BibitemOpen
  \bibfield  {author} {\bibinfo {author} {\bibfnamefont {T.}~\bibnamefont
  {Hahn}}\ and\ \bibinfo {author} {\bibfnamefont {M.}~\bibnamefont
  {Perez-Victoria}},\ }\href {\doibase 10.1016/S0010-4655(98)00173-8}
  {\bibfield  {journal} {\bibinfo  {journal} {Comput. Phys. Commun.}\ }\textbf
  {\bibinfo {volume} {118}},\ \bibinfo {pages} {153} (\bibinfo {year}
  {1999})},\ \Eprint {http://arxiv.org/abs/hep-ph/9807565}
  {arXiv:hep-ph/9807565 [hep-ph]} \BibitemShut {NoStop}%
\bibitem [{\citenamefont {Shtabovenko}\ \emph {et~al.}(2016)\citenamefont
  {Shtabovenko}, \citenamefont {Mertig},\ and\ \citenamefont
  {Orellana}}]{Shtabovenko:2016sxi}%
  \BibitemOpen
  \bibfield  {author} {\bibinfo {author} {\bibfnamefont {V.}~\bibnamefont
  {Shtabovenko}}, \bibinfo {author} {\bibfnamefont {R.}~\bibnamefont {Mertig}},
  \ and\ \bibinfo {author} {\bibfnamefont {F.}~\bibnamefont {Orellana}},\
  }\href {\doibase 10.1016/j.cpc.2016.06.008} {\bibfield  {journal} {\bibinfo
  {journal} {Comput. Phys. Commun.}\ }\textbf {\bibinfo {volume} {207}},\
  \bibinfo {pages} {432} (\bibinfo {year} {2016})},\ \Eprint
  {http://arxiv.org/abs/1601.01167} {arXiv:1601.01167 [hep-ph]} \BibitemShut
  {NoStop}%
\bibitem [{\citenamefont {Patel}(2015)}]{Patel:2015tea}%
  \BibitemOpen
  \bibfield  {author} {\bibinfo {author} {\bibfnamefont {H.~H.}\ \bibnamefont
  {Patel}},\ }\href {\doibase 10.1016/j.cpc.2015.08.017} {\bibfield  {journal}
  {\bibinfo  {journal} {Comput. Phys. Commun.}\ }\textbf {\bibinfo {volume}
  {197}},\ \bibinfo {pages} {276} (\bibinfo {year} {2015})},\ \Eprint
  {http://arxiv.org/abs/1503.01469} {arXiv:1503.01469 [hep-ph]} \BibitemShut
  {NoStop}%
\bibitem [{\citenamefont {Kinoshita}(1962)}]{Kinoshita:1962ur}%
  \BibitemOpen
  \bibfield  {author} {\bibinfo {author} {\bibfnamefont {T.}~\bibnamefont
  {Kinoshita}},\ }\href {\doibase 10.1063/1.1724268} {\bibfield  {journal}
  {\bibinfo  {journal} {J. Math. Phys.}\ }\textbf {\bibinfo {volume} {3}},\
  \bibinfo {pages} {650} (\bibinfo {year} {1962})}\BibitemShut {NoStop}%
\bibitem [{\citenamefont {Lee}\ and\ \citenamefont
  {Nauenberg}(1964)}]{Lee:1964is}%
  \BibitemOpen
  \bibfield  {author} {\bibinfo {author} {\bibfnamefont {T.~D.}\ \bibnamefont
  {Lee}}\ and\ \bibinfo {author} {\bibfnamefont {M.}~\bibnamefont
  {Nauenberg}},\ }\href {\doibase 10.1103/PhysRev.133.B1549} {\bibfield
  {journal} {\bibinfo  {journal} {Phys. Rev.}\ }\textbf {\bibinfo {volume}
  {133}},\ \bibinfo {pages} {B1549} (\bibinfo {year} {1964})}\BibitemShut
  {NoStop}%
\bibitem [{\citenamefont {Minkowski}(1977)}]{Minkowski:1977sc}%
  \BibitemOpen
  \bibfield  {author} {\bibinfo {author} {\bibfnamefont {P.}~\bibnamefont
  {Minkowski}},\ }\href {\doibase 10.1016/0370-2693(77)90435-X} {\bibfield
  {journal} {\bibinfo  {journal} {Phys. Lett.}\ }\textbf {\bibinfo {volume}
  {67B}},\ \bibinfo {pages} {421} (\bibinfo {year} {1977})}\BibitemShut
  {NoStop}%
\bibitem [{\citenamefont {Gell-Mann}\ \emph {et~al.}(1979)\citenamefont
  {Gell-Mann}, \citenamefont {Ramond},\ and\ \citenamefont
  {Slansky}}]{GellMann:1980vs}%
  \BibitemOpen
  \bibfield  {author} {\bibinfo {author} {\bibfnamefont {M.}~\bibnamefont
  {Gell-Mann}}, \bibinfo {author} {\bibfnamefont {P.}~\bibnamefont {Ramond}}, \
  and\ \bibinfo {author} {\bibfnamefont {R.}~\bibnamefont {Slansky}},\
  }\bibfield  {booktitle} {\emph {\bibinfo {booktitle} {{Supergravity Workshop
  Stony Brook, New York, September 27-28, 1979}}},\ }\href@noop {} {\bibfield
  {journal} {\bibinfo  {journal} {Conf. Proc.}\ }\textbf {\bibinfo {volume}
  {C790927}},\ \bibinfo {pages} {315} (\bibinfo {year} {1979})},\ \Eprint
  {http://arxiv.org/abs/1306.4669} {arXiv:1306.4669 [hep-th]} \BibitemShut
  {NoStop}%
\bibitem [{\citenamefont {Yanagida}(1979)}]{Yanagida:1979as}%
  \BibitemOpen
  \bibfield  {author} {\bibinfo {author} {\bibfnamefont {T.}~\bibnamefont
  {Yanagida}},\ }\bibfield  {booktitle} {\emph {\bibinfo {booktitle}
  {{Proceedings: Workshop on the Unified Theories and the Baryon Number in the
  Universe: Tsukuba, Japan, February 13-14, 1979}}},\ }\href@noop {} {\bibfield
   {journal} {\bibinfo  {journal} {Conf. Proc.}\ }\textbf {\bibinfo {volume}
  {C7902131}},\ \bibinfo {pages} {95} (\bibinfo {year} {1979})}\BibitemShut
  {NoStop}%
\bibitem [{\citenamefont {Mohapatra}\ and\ \citenamefont
  {Senjanovic}(1980)}]{Mohapatra:1979ia}%
  \BibitemOpen
  \bibfield  {author} {\bibinfo {author} {\bibfnamefont {R.~N.}\ \bibnamefont
  {Mohapatra}}\ and\ \bibinfo {author} {\bibfnamefont {G.}~\bibnamefont
  {Senjanovic}},\ }\href {\doibase 10.1103/PhysRevLett.44.912} {\bibfield
  {journal} {\bibinfo  {journal} {Phys. Rev. Lett.}\ }\textbf {\bibinfo
  {volume} {44}},\ \bibinfo {pages} {912} (\bibinfo {year} {1980})},\ \bibinfo
  {note} {[,231(1979)]}\BibitemShut {NoStop}%
\end{thebibliography}%
\end{document}